\documentclass[]{article}
\usepackage[]{amsmath}
\usepackage{amssymb}

\numberwithin{equation}{section}
\newtheorem{theorem}[equation]{Theorem}
\newtheorem{corollary}[equation]{Corollary}
\newtheorem{lemma}[equation]{Lemma}
\newtheorem{proposition}[equation]{Proposition}
\newtheorem{definition}[equation]{Definition}
\begin{document}

\title{Mutually Unbiased Bases, Generalized Spin Matrices and 
Separability}
\author{Arthur O. Pittenger\\ 
Department of Mathematics and Statistics\\
\and 
Morton H. Rubin\\
Department of Physics\\
\\
University of Maryland, Baltimore County\\
 Baltimore, MD 21250\\
 pittenge@math.umbc.edu\\
rubin@umbc.edu}
\date{August 25, 2003}

\maketitle

\begin{abstract}
A collection of orthonormal bases for a complex $d$-dimensional Hilbert space is called 
{\it mutually unbiased} (MUB) if for any two vectors $v$ and $w$ from
different bases the square of the inner product equals $1/d$: 
$\left| \left\langle v,w\right\rangle \right| ^{2}=\frac{1}{d}$. The MUB problem is
to prove or disprove the existence of a maximal set of $d+1$ bases.
It has been shown in W. K. Wootters and B. D. Fields (1989, Annals of Physics, 
 {\bf 191}, 363) that such a collection exists if $d$
is a power of a prime number $p$. We revisit this problem and use $d\times d$
generalizations of the Pauli spin matrices to give a constructive proof of
this result. Specifically we give explicit representations of commuting
families of unitary matrices whose eigenvectors solve the MUB problem.
Additionally we give formulas from which the orthogonal bases can be readily
computed. We show how the techniques developed here provide a natural 
way to analyze the separability of the bases. The techniques used require properties of algebraic field
extensions, and the relevant part of that theory is included in an
Appendix.
\end{abstract}



\section{Introduction.}

Let $H$ denote a complex $d$-dimensional Hilbert space and $\rho $ a density matrix
modeling a $d$-level quantum system. Then $\rho $ is a positive
semidefinite, trace one matrix and as such is Hermitian and is determined by 
$d^{2}-1$ real numbers. A laboratory device that measures $\rho $ is
represented by a Hermitian matrix $A=\sum_{k=1}^{d}\lambda _{k}P_{k}$, where 
$\left\{ P_{k}\text{:}1\leq k\leq n\right\} $ is a set of rank one mutually
orthogonal projections. (In Dirac notation $P_{k}$ denotes the outer product 
$\left| v_{k}\right\rangle \left\langle v_{k}\right| $ of the eigenvector $%
\left| v_{k}\right\rangle $.)  If the eigenvalues are distinct, $A$ is
called non-degenerate, and the non-negative values $p_{k}\left( \rho
,A\right) =Tr\left[ \rho P_{k}\right] $ can be estimated by repeated
experiments. Since  $\sum_{k}p_{k}\left( \rho, A\right) =1$, one 
obtains $d-1$ independent pieces of information, and a minimum of $d+1$ 
such well designed experiments would be required to recover the density $\rho$.

The problem of mutually unbiased bases (MUB) refers to the theoretical
possibility of defining $d+1$ such bases with the additional property that $%
Tr\left( P_{j}^{r}P_{k}^{s}\right) =\frac{1}{d}$ for any pair of projections
associated with different experimental configurations, labeled by $r$ 
and $s$. Such a collection of
bases provides an optimal way of estimating $\rho $, and we refer to \cite
{Woot} for a discussion of that feature.

As an example, for a two-level system there is such a set of bases that can be
represented in terms of the usual Pauli matrices, 
\begin{displaymath}
	\sigma_{0}=\left(
	\begin{array} {cc}
	1 & 0 \\
	0&1\\
	\end{array}\right)
	\sigma_{x}=\left(
	\begin{array} {cc}
	0 & 1 \\
	1&0\\
	\end{array}\right)\sigma_{y}=\left(
	\begin{array} {cc}
	0 & -i \\
	i&0\\
	\end{array}
	\right)\sigma_{z}=\left(
	\begin{array} {cc}
	1 & 0 \\
	0&-1\\
	\end{array}\right).
\end{displaymath}
 The three sets of
projections $\left\{ \frac{1}{2}\left( \sigma _{0}\pm \sigma _{x}\right)
\right\} $, $\left\{ \frac{1}{2}\left( \sigma _{0}\pm \sigma _{y}\right)
\right\} $, and $\left\{ \frac{1}{2}\left( \sigma _{0}\pm \sigma _{z}\right)
\right\} $ correspond to measurements along the three spin axes of a two-level
system. The existence of such bases for $d=p$, $p$ a prime, was first
established in \cite{IDI} and was extended to $d=p^{n}$ in \cite{Woot}.
Recent papers on the subject include \cite{BBRV,klap}, that discuss the
general case, and \cite{zeil}, that works in the context of $d=2^{n}$. To
the best of our knowledge, there are no definitive results for other values
of $d$.

While writing up our results, we attended a talk by
Bill Wootters, who outlined a different approach to the problem of mutually
unbiased bases and who brought \cite{Woot2} to our attention. Although the
motivations of the two approaches appear to be quite different, they require
the same mathematical tools and appear to lead to the same results. An
interesting question is the relationship between the two approaches. 

Our interest in this problem was stimulated by the following result in \cite
{BBRV}.
\begin{theorem}
(\cite{BBRV} Thm 3.2) Suppose that one has $d^{2}$ unitary matrices
orthogonal in the Frobenius or trace inner product, one of which is the identity matrix.
Suppose further that these matrices can be grouped into $d+1$ classes of $d$
commuting matrices and that the only matrix common to two different classes
is the identity. Then there is a set of $d+1$ mutually unbiased bases.
\label{th1}
\end{theorem}

Motivated by the observation that the Pauli spin matrices can be derived as
a Hadamard transform of certain basis matrices, we defined in \cite{PR1} a
family of $d^{2}$ matrices that are orthogonal with respect to the trace
inner product. Accordingly we refer to them as (generalized) spin matrices.
Although that approach seems to have been novel, these matrices have
appeared earlier in the literature, for example in \cite{cal} and \cite{Fiv}
and references therein. They were also used in \cite{BBRV}.

In addition to providing an algorithm for deriving explicit solutions to the
MUB problem for $d=p^{n}$, a major goal of this paper is to emphasize the
utility of the indexing of the generalized spin matrices. In fact, by
interpreting the indices as vectors we are able to put the MUB problem into
the context of a vector space over a finite field. Moreover, we can also use
the indexing and results in \cite{PR1} to write each mutually unbiased basis
defined by a set of commuting matrices as a weighted sum of those matrices.

In Section 2 we define the generalized spin matrices and record a number of
the properties given in \cite{PR1}. In Section 3 we use the notation of the
generalized spin matrices to facilitate a detailed solution of the mutually
unbiased bases problem when $d=p$ is an odd prime. A basic idea
used in that solution reappears in the next two sections. In Section 4 we
show how the use of (algebraic) field extensions produces a solution for $%
d=p^{2}$ and set the stage for Section 5, in which we give a constructive
algorithm for solving the MUB problem explicitly in the general case of $%
d=p^{n}$. In Section 6 we define the notion of separability of a basis and
show how the separability of the derived bases is related to the index
notation. To improve the readability of the paper, we have deferred many of
the technicalities to the end of the paper. Thus the Appendices provide the
details for computing the projections associated with a class of commuting
spin matrices, the formal mathematics underlying the results in Section 4,
and the theoretical foundation for the algorithm illustrated in Section 5.

It is important to emphasize that our methodology gives a specific solution
of the MUB problem for $d=p^{n}$. Once such a solution is in hand, there are
many ways to construct other mutually unbiased bases, such as using
conjugation by a unitary matrix. 

Finally a word about notation. Throughout the paper we use the letters 
$j$, $k$, $a$, $b$ to denote the elements of $Z_{d},$ the integers modulo $d$.
 The letters $u$, $v$, and $z$ denote vectors in $V_{2}\left( F\right) $, the
two dimensional vector space over a field $F$, and $w$ denotes a vector
in $V_{2n}\left( Z_{p}\right) $, the $2n$-dimensional vector space over $%
Z_{p}$, where $p$ is a prime. The Greek letters $\alpha $, $\beta $ are
reserved for elements of the Galois field $GF\left( p^{n}\right) $. 

\section{Generalized spin matrices}

In what follows $d$ denotes the dimension of the finite dimensional 
complexHilbert space $H$, and the unitary matrices acting on $H$ are indexed by subscripts 
$u=\left(j,k\right) $, with the two forms of indices used interchangeably. 
Let $\{|j\rangle, j=0,\cdots,d-1\}$ be a fixed orthonormal basis of $H$.
We will have occasion to use vector addition of indices, and such addition will be
addition modulo $d$. $\eta $ denotes the complex number $\exp \left( 2\pi
i/d\right) $, and it is easy to confirm that for integers $b$ such that $
\eta ^{b}\neq 1$

\[
\sum_{k=0}^{d-1}\left( \eta ^{b}\right) ^{k}=0. 
\]

\begin{definition}
Let $0\leq j,k<d$. Then $S_{j,k}\equiv $ $\sum_{m=0}^{d-1}\eta ^{mj}\left|
m\right\rangle \left\langle m+k\right| $.
\label{defS}
\end{definition}

It is easy to confirm that $Tr(S_{j,k})=0$ unless $S_{j,k}=S_{0,0},$ the $%
d\times d$ identity matrix. A key property is that this set of matrices is
closed under multiplication, up to scalar multiples of powers of $\eta $.

\begin{lemma}
$S_{j,k}S_{a,b}=\eta ^{ka}S_{j+a,k+b}.$ Thus, $S_{j,k}$ and $S_{a,b}$
commute if and only if $ka=jb$ up to an additive multiple of $d.$
\label{lemmaProd}
\end{lemma}
{\it Proof}:  Using the obvious notation, 
\[
S_{j,k} S_{a,b}=\sum_{m=0}^{d-1}\sum_{n=0}^{d-1}\eta ^{mj+na}
\delta(m+k,n) \left| m\right\rangle
\left\langle n+b\right|.
\]
If $m+k\leq d-1,$ $n=m+k$ gives the only non-zero factor. If $m+k\geq d$, $%
n=m+k-d$ gives the only non-zero factor. Since $\eta ^{d}=1$, we have $%
S_{j,k} S_{a,b}=\eta ^{ka}\sum_{m=0}^{d-1}\eta ^{m\left( j+a\right)
}\left| m\right\rangle \left\langle m+k+b\right|. \quad  \Box$

Some useful relations follow immediately, with ({\it iii}) established by
induction. ($S_{0,1}$ and $S_{1,0}$ are generators of the set 
$\{S_{j,k}\}$ and reduce to $\sigma_{x}$ and 
$\sigma_{z}$ when $d=2$.)

\begin{corollary}
 ({\it i}) $S_{0,1} S_{1,0}=\eta S_{1,1}=\eta S_{1,0} 
 S_{0,1}$, 
({\it ii}) $S_{j,k}=\left( S_{1,0}\right) ^{j}\left( S_{0,1}\right) 
^{k},$\\
({\it iii}) 
\begin{equation}
	\left( S_{j,k}\right) ^{m}=\eta ^{jk\binom{m}{2}}S_{mj,mk}
\end{equation}
where $\binom{m}{2} \equiv 0$ for $m=0$ or $1$.
\label{corSpower}
\end{corollary}

We next establish that these matrices are unitary and are also orthogonal to
one another with respect to the Frobenius inner product on the space of $%
d\times d$ complex matrices, $\langle A,B\rangle =tr(A^{\dagger}B)$, 
where $A^{\dagger}$ is the Hermitian conjugate of $A$.

\begin{lemma}
$\left( S_{j,k}\right) ^{\dagger }=\eta ^{jk}S_{-j,-k}$. For each $u$, $S_{u}
$ is unitary, and \\
 $Tr\left[ \left( S_{u}\right) ^{\dagger }S_{v}\right] =0$
if $u\neq v$.
\end{lemma}
{\it Proof}: \[
\left( S_{j,k}\right) ^{\dagger }=\sum_{m=0}^{d-1}\eta ^{-mj}\left|
m+k\right\rangle \left\langle m\right| =\eta ^{jk}\sum_{n=k}^{d+k-1}\eta
^{-nj}\left| n\right\rangle \left\langle n-k\right| =\eta ^{jk}S_{-j,-k.}
\]
 Let $u=(j,k), v=\left( a,b\right) $; then
 \[
\left( S_{u}\right) ^{\dagger }S_{v}=\eta ^{jk}S_{-j,-k}S_{a,b}=\eta
^{k\left( j-a\right) }S_{a-j,b-k}. 
\]
This has trace zero if $u\neq v,$ and if $u=v$, we get the identity, so that $%
S_{u}$ is unitary. $\quad \Box$

It follows that $\left\{ S_{u}:u=\left( j,k\right) \right\} $ is a set of $%
d^{2}$ unitary matrices that forms an orthogonal basis for the space
of $d\times d$ matrices and is closed under multiplication, up to
multiples of powers of $\eta $. Thus they can be regarded as analogues of
the Pauli spin matrices, hence the terminology \textit{generalized spin 
matrices}.

One doesn't quite recover the Pauli matrices through this procedure. In fact
when $d=2$, one has $S_{0,1}=\sigma _{x},$ $S_{1,0}=\sigma _{z},$ but $
S_{1,1}=i\sigma _{y}$ in order to fit into the general framework. The missing factor of $i=\left( -1\right) ^{1/2}$
reappears when we define the projections associated with these unitary
matrices.

Such orthogonal families of unitary matrices play a key role in quantum
information theory, as elaborated in \cite{Wer}, and, as established in
Theorem \ref{th1}, they are closely related to solutions of the MUB problem. The
proof of Theorem \ref{th1} uses the fact that commuting unitary matrices can be
simultaneously diagonalized, and the bases related to the different
classes have the MUB property. The orthogonality of the unitary
matrices is crucial to the analysis, and thus the connection to the
generalized spin matrices is immediate. Our problem then reduces to finding
commuting classes, and the characterization of commutativity in terms of the
indices enables us to rephrase the problem as a vector space problem over a
finite (algebraic) field. By using this specific class of orthogonal unitary
matrices, we are also able to give explicit formulas for the projections
defined by the basis vectors.

\section{Spin matrices and the MUB problem for $d$ prime}

We begin with the case when $d=p$ is a prime. As we have seen, $%
S_{j,k}$ and $S_{a,b}$ commute if and only if $ka=jb \bmod p$. We recast 
this condition in the context of a vector
space over the finite field $Z_{p}$, the integers modulo the prime $p$. Let 
$V_{2}(Z_{p}) =\left\{ (j,k):j,k\in Z_{p}\right\} $, and define a {\it symplectic}
product: 
\begin{equation}
u\circ u^{\prime}\equiv kj^{\prime}-jk^{\prime} \quad\bmod p  
\label{symprod}
\end{equation}
where  $u=\left( j,k\right) $ and $u^{\prime}=\left( j^{\prime},k^{\prime}\right)$.
Thus, $S_{u}$ and $S_{v}$ commute if and only if the symplectic product of
their vector indices equals zero.

Once we have the classes of commuting matrices, we can make a direct 
computation (or invoke Theorem 1.1) to
argue the existence of a complete set of mutually unbiased bases. We 
can construct these bases explicitly in terms of the spin matrices as 
follows.
\begin{proposition}
Let $a\in Z_{p}$ and define
 \begin{eqnarray*}
C_{a}&=&\left\{ b \left( 1,0\right) +ba\left( 0,1\right) 
=b\left( 1,a\right) :b\in Z_{p}\right\} \\
 C_{\infty}&=&\left\{ b\left(
0,1\right) :b\in Z_{p}\right\}.
\end{eqnarray*}
There are $p$ vectors in each of
these $p+1$ classes and $C_{r}\cap C_{s}=\left\{ \left( 0,0\right) \right\} $
for all $r\neq s$ in $I\equiv \left\{ 0,1,\ldots ,p-1,\infty \right\} .$ If $%
u,v$ are in $C_{r}$, then $u\circ v=0.$
\label{propClasses}
\end{proposition}
{\it Proof}: The vectors $e=\left( 1,0\right) $ and $f=\left( 0,1\right) $
are linearly independent with $f\circ e=1$ and $e\circ e=f\circ f=0.$ If $%
b\left( 1,a\right) =b^{^{\prime} }\left( 1,a^{^{\prime} }\right) $, then $
b=b^{^{\prime} }$ and if $b\neq 0$, $a=a^{^{\prime} }.$ This proves the first
assertion for the $C_{a}$ classes. Using the linearity of the symplectic
product, 
\[
\left[ b\left( 1,a\right) \right] \circ \left[ c\left( 1,a\right) \right]
=bc\left( 1,a\right) \circ \left( 1,a\right) =0. 
\]
The same arguments work for $C_{\infty }. \quad \Box $\\
 
 The $C_{t}$ can be thought of as lines in a two-dimensional space. 
In addition the vectors in $C_{t}$ can be written as a multiple of a single
vector $u_{t}=\left( j_{t},k_{t}\right) $, and $C_{t}$ is an additive
subgroup of $V_{2}(Z_{p})$. The matrices associated with $C_{t}$
are $\left\{ S_{nu_{t}},0\leq n<p\right\} $; they commute but do not
form a multiplicative subgroup of the unitary matrices by virtue of
Corollary \ref{corSpower} ({\it iii}). We nonetheless consider $S_{u_{t}}$ to be
the ``generator'' of  $\left\{ S_{nu_{t}},0\leq n<p\right\} $ with the
understanding that it is $S_{nu_{t}}$, not $\left( S_{u_{t}}\right) ^{n}=\eta
^{j_{t}k_{t}\binom{n}{2}}S_{nu_{t}}$ that is in the class.  

Theorem \ref{th1} guarantees that the orthonormal eigenvectors for each class
solve the MUB problem, and we can use the indicial notation to express the
associated orthogonal projections explicitly in terms of the unitary
matrices \cite{PR1}. We begin with a definition that is valid for all $d$
and is required to handle the computations in general.

\begin{definition}
Let $0 \leq j,k < d$ and $u=(j,k)$. If $d$ is even and both $j$ and $k$ are odd, 
set $\alpha _{u}=-\exp \left( \pi i/d\right) =-\eta ^{1/2}$.
Otherwise set $\alpha _{u}=1$.
\label{defAlpha}
\end{definition}

For example, for $d=2$ and $j=k=1$, $\alpha _{u}=-i$. In general, for $d\geq 2$, $\alpha _{u}^{d}\eta
^{jk\binom{d}{2}}=1.$

\begin{definition}
For each $u=(j,k)\neq \left( 0,0\right) $ and $0\leq r<d,$ define 
\begin{equation}
P_{u}\left( r\right) =\frac{1}{d}\sum_{m=0}^{d-1}\left( \alpha _{u}\eta
^{r}S_{u}\right) ^{m},  \label{proj1}
\end{equation}
where $\left( \alpha _{u}\eta ^{r}S_{u}\right) ^{0}\equiv S_{0,0}$.
\label{propProj1}
\end{definition}
\begin{proposition}
 For $d$ a prime, $\left\{ P_{u}\left( r\right) :0\leq r<d\right\} $ is a complete set of
mutually orthogonal projections.
\label{propProj}
\end{proposition}
It is easy to check that $P_{u}\left( r\right) $ has trace one and that 
\begin{equation}
\left( \alpha _{u}\eta ^{r}S_{u}\right) ^{t}=\sum_{m=0}^{d-1}\eta
^{-mt}P_{u}\left( m+r\right) ,  \label{spin/proj}
\end{equation}
(\cite{PR1}, equation (13)). We need to confirm that the $P_{u}\left(
r\right) $'s constitute a set of $d$ orthogonal, one-dimensional projections, and we
provide the details in Appendix A.

As just noted, the indices of members of a commuting class are multiples of
a vector $u_{t}$. Thus if $u=bu_{t}$, then $P_{u}\left( r\right) $ should be 
$P_{u_{t}}\left( s\right) $ for some $s$, and we confirm that fact next.

\begin{corollary}
If $p>2$ is prime and $u=bu_{t}=b\left( j_{t},k_{t}\right) $ with $2\leq b<p$%
, then $P_{u}\left( r\right) =P_{u_{t}}\left( s\right)$, where $
s=b^{-1}\left( r-j_{t}k_{t}\binom{b}{2}\right) $ and $b^{-1}$ is the
multiplicative inverse of $b$ modulo $p$.
\end{corollary}
{\it Proof}: From (iii)\ (\ref{corSpower}), it follows that $\left( S_{u}\right) ^{m}=\eta ^{-mj_{t}k_{t}\binom{b}{2}%
}\left( S_{u_{t}}\right) ^{bm}.$ Hence 
\[
P_{u}\left( r\right) =\frac{1}{d}\sum_{m=0}^{d-1}\eta ^{m\left( r-j_{t}k_{t}%
\binom{b}{2}\right) }S_{u_{t}}^{bm}=\frac{1}{d}\sum_{n=0}^{d-1}\eta
^{nb^{-1}\left( r-j_{t}k_{t}\binom{b}{2}\right)
}S_{u_{t}}^{n}=P_{u_{t}}\left( s\right) ,
\]
where we made the substitution $n=bm \bmod p. \quad \Box$

We now show that $Tr\left[ P_{u}\left( r\right) P_{u^{\prime}}\left(
s\right) \right] =1/d$, where it suffices to take $u=\left( 
1,a\right) $ and $u^{\prime}=\left(
1,a^{\prime}\right) $ as representatives of different classes $C_{a}$. In general
\[
P_{u}\left( r\right) P_{u^{\prime}}\left( s\right) =\frac{1}{p^{2}}%
\sum_{m=0}^{p-1}\sum_{n=0}^{p-1}\alpha _{u}^{m}\eta ^{mr+a\binom{m}{2}%
}\alpha _{u^{\prime}}^{n}\eta ^{ns+a^{\prime}\binom{n}{2}}S_{mu}S_{nu^{\prime}},
\]
and we see that the only contribution to the trace is for 
$mu+nu^{\prime}=(0,0) \bmod p$. (Again, $\binom{m}{2}$ is taken to be zero if $m$ $=0$
or $1$.) This means that $m$ and $n$ satisfy 
\[
\left( 
\begin{array}{cc}
1 & 1 \\ 
a & a^{\prime}
\end{array}
\right) \left( 
\begin{array}{c}
m \\ 
n
\end{array}
\right) =\left( 
\begin{array}{c}
0 \\ 
0
\end{array}
\right)  \bmod p.
\]
Since $a\neq a^{\prime}$, only $m=n=0$ satisfy the equation. Hence $Tr\left[
P_{bu}\left( r\right) P_{b^{\prime}u^{\prime}}\left( s\right) \right] 
=1/d$ as required.
The details for $C_{\infty }$ are similar. We now have proved the following 
theorem that recaptures the basic result of \cite{IDI}.
\begin{theorem}
If $p$ is prime, there is a complete set of $p+1$ mutually unbiased bases $%
B_{a}$, $0\leq a<p$, and $B_{\infty }$ that are the normalized eigenvectors of the
corresponding sets of commuting spin matrices $\left\{
S_{b,ba}:b\in Z_{p}\right\}\leftrightarrow C_{a}$ and $\left\{
S_{0,b}:b\in Z_{p}\right\}\leftrightarrow C_{\infty }$. These bases can be computed from the projections 
in eq.\ (\ref{proj1}).
\end{theorem}

\textit{Example}: The classes for $d=2$ are $\left\{ S_{0,0},S_{1,0}\right\} $, 
$\left\{ S_{0,0},S_{1,1}\right\} $, and $\left\{ S_{0,0},S_{0,1}\right\} $,
where $S_{1,0}=\sigma _{z}$, $S_{0,1}=\sigma _{x},$ and $S_{1,1}=i\sigma
_{y} $. The MUB's are determined by the projectors $\frac{1}{2}\left( \sigma _{0}\pm \sigma _{z}\right) 
$, $\frac{1}{2}\left( \sigma _{0}\pm \sigma _{y}\right) $, and $\frac{1}{2}%
\left( \sigma _{0}\pm \sigma _{x}\right) $ from (\ref{proj1}).
The factor $\alpha _{1,1}=-i$ is needed to recover the projections $%
\frac{1}{2}\left( \sigma _{0}\pm \sigma _{y}\right) $ from the general
formula.

We obtain four classes of commuting spin matrices for $d=3$ and can
represent them in a $3\times 3$ table, where the row index denotes $j$ and
the column index $k$ in $S_{j,k}.$ Similar tables can be constructed for
larger values of $p$, and in a finite geometry interpretation the 
classes $C_{r}$ determine lines intersecting only at the origin.  
\[
\begin{array}{cccc}
& 0 & 1 & 2 \\ 
0 &  & C_{\infty } & C_{\infty } \\ 
1 & C_{0} & C_{1} & C_{2} \\ 
2 & C_{0} & C_{2} & C_{1}
\end{array}
\]

An additional feature of the spin matrices allows one to express estimates
of the components of a density $\rho $ in the original fixed basis in terms
of measurements in the MUB bases. We sketch the idea. Assume $d=p$ and
express the density matrix as 
\[
\rho =\frac{1}{p}\left[ \sum_{j,k=0}^{p-1}s_{j,k}S_{j,k}\right] =\frac{1}{p}%
\left[ S_{00}+\sum_{t\in I}\sum_{u\in C_{t}-\left\{ \left( 0,0\right)
\right\} }s_{u}S_{u}\right] ,
\]
where $I\equiv \left\{ 0,1,\ldots ,p-1,\infty \right\} $. From the
orthogonality of the spin matrices and their representation in terms of the
projections of their commuting class, we know that 
\begin{equation}
s_{u}=Tr\left( S_{u}^{\dagger }{}\rho \right) =\alpha
_{u}\sum_{m=0}^{d-1}\eta ^{m}p_{u}\left( m\right),
\label{spincoeff}
\end{equation}
where $p_{u}\left( m\right) \equiv Tr\left( P_{u}\left( m\right) \rho
\right)$. 
A measuring device $M_{u}$ may be characterized by 
 $\{P_{u}(m), 0 \leq 
m < p\}.$  If the system is in a state modeled by the density 
$\rho$, $M_{u}$ determines the probability, $p_{u}(m)$, of the outcome $m$. The 
experimental results of measurements over an ensemble of systems give estimates
for these probabilities and, by (\ref{spincoeff}), estimates for all of the spin
coefficients with indices in that commuting class. Since the spin
coefficients themselves are Fourier transforms of entries of $%
\rho $ in the original basis (\cite{PR1}, equation (11)), it follows
that an estimate of $\rho$ in this basis can be expressed
explicitly in terms of measurements in the MUB bases. For a more 
complete discussion of the estimation problem see \cite{Woot}.

\section{The MUB problem for $d=p^{2}$, $p$ an odd prime}

It was shown in \cite{Woot} that the MUB problem can be solved for powers of
primes. We give a concrete construction based
on algebraic techniques and motivated by the results in the preceding
section and Theorem 1.1. This requires a certain amount of abstract
algebra, and we present the special case of $d=p^{2}$ to illustrate
the results and the ideas. (The case $p=2$ requires a modification of the
approach used here and is discussed in the next section.) However, the basic
strategy is the same as before. We use the indices of the spin matrices to
encode commutativity and techniques of vector spaces over finite fields to define the
appropriate classes. The actual MUB bases can then be recovered from the
classes of commuting spin matrices.

We are working with tensor products of the form $S_{u}\otimes S_{v}$, 
where commutativity is again encoded by the indices
so that $S_{u_{1}}\otimes S_{v_{1}}$ commutes with $S_{u_{2}}\otimes S_{v_{2}}$ if and only if 
\[
u_{1} \circ u_{2}
+v_{1} \circ  v_{2}=0 \bmod p,
\]
where $u=\left(j,k\right)$ and $v=(a,b)$.
It is now useful to consider vectors in a
four dimensional vector space over $Z_{p}$,
$V_{4}(Z_{p})=\{w=(j,k,a,b)=(u,v)\}$, and
to define the symplectic product on the four dimensional space as
\begin{equation}
w_{1}\circ w_{2}\equiv u_{1}\circ u_{2}+v_{1}\circ v_{2}.  \label{symprod1}
\end{equation}
The first two indices in $w$ correspond to the indices in the
first factor and the second two indices correspond to
the second factor in the tensor product $S_{u}\otimes S_{v}$.

The solution to the problem of finding the commuting classes of spin 
matrices now reduces to finding the classes of vectors $w$ that satisfy $w_{1}\circ 
w_{2}=0$.  A technology for doing this is 
discussed in Appendix C.  Here we simply give the results.

For $p$ an odd prime, the procedure to define classes of
four-vectors with symplectic products equal to zero requires a particular
non-zero integer $D$ in $Z_{p}$. $D$ is defined by the 
requirement that $D\neq k^{2} \bmod p$ for all $k$ in $Z_{p}$, 
\textit{i.e.} $D$ is not a quadratic residue of $p$.

\begin{theorem}
Let $p$ be an odd prime. Then commuting classes of spin matrices are indexed by
the following subsets of $V_{4}\left( Z_{p}\right):$%
\[
C_{a_{0},a_{1}}=\left\{ \left(
2b_{0},a_{0}b_{0}+a_{1}b_{1}D,2b_{1}D,a_{0}b_{1}+a_{1}b_{0}\right)
:b_{0},b_{1}\in Z_{p}\right\} 
\]
\[
C_{\infty }=\left\{ \left( 0,b_{0},0,b_{1}\right) :b_{0},b_{1}\in
Z_{p}\right\},
\]
where $a_{0},a_{1}\in Z_{p}$ and $\left( j_{1},k_{1},j_{2},k_{2}\right) $ corresponds to $%
S_{j_{1},k_{1}}\otimes S_{j_{2},k_{2}}$. $C_{a_{0},a_{1}}$ is a 
subspace of $V_{4}(Z_{p})$ with basis 
\[
G_{a_{0},a_{1}} = \{(2,a_{0},0,a_{1}), (0,a_{1}D,2D,a_{0})\}
\]
and $C_{\infty}$ has the basis $G_{\infty}=\{(0,1,0,0), (0,0,0,1) \}$.
\label{thN2Classes}
\end{theorem}

The structure of $C_{a_{0},a_{1}}$ is hardly an intuitive result, but 
we take it as given and confirm
the desired properties. There are $p^{2}+1$ such classes. We claim that each class
has $p^{2}$ members, that $w_{1}\circ w_{2}=0$ for vectors in the same
class, and that the only vector common to any pair of classes is $\left(
0,0,0,0\right) $. If so, then the classes partition $%
V_{4}\left( Z_{p}\right) -\left\{ \left( 0,0,0,0\right) \right\} $ as
required.

The verification of these three properties is quite easy, and we leave the
details to the reader. We should note, however, that in checking the last
property we are led to the equations 
\begin{eqnarray*}
a_{0}b_{0}+a_{1}b_{1}D &=&a_{0}^{^{\prime} }b_{0}+a_{1}^{^{\prime} }b_{1}D \\
a_{0}b_{1}+a_{1}b_{0} &=&a_{0}^{^{\prime} }b_{1}+a_{1}^{^{\prime} }b_{0},
\end{eqnarray*}
where $a_{0}$, $a_{1}$ and $a_{0}^{^{\prime} }$, $a_{1}^{^{\prime} }$ denote
indices of the first type of class and $b_{0}\neq 0\neq b_{1}$. This system
can be rewritten as a matrix equation 
\[
\left( 
\begin{array}{cc}
b_{0} & b_{1}D \\ 
b_{1} & b_{0}
\end{array}
\right) \left( 
\begin{array}{c}
a_{0}-a_{0}^{^{\prime} } \\ 
a_{1}-a_{1}^{^{\prime} }
\end{array}
\right) =\left( 
\begin{array}{c}
0 \\ 
0
\end{array}
\right) 
\]
that has only the trivial solution provided $b_{1}^{2}D\neq b_{0}^{2} 
\bmod{p}$.
Since  $x^{2}=D$ is {\it not} solvable in $Z_{p}$,
all of the properties hold and we have classes of commuting spin
matrices of the form $S_{2b_{0},a_{0}b_{0}+a_{1}b_{1}D}\otimes
S_{2b_{1}D,a_{0}b_{1}+a_{1}b_{0}}$ indexed by $a_{0}$ and $a_{1}.$ The
matrices associated with $C_{\infty }$ have the form $S_{0,b_{0}}\otimes
S_{0,b_{1}}$.

We can always find such values $D$. For example, if $p=3$, $D=2$%
; if $p=5$, $D$ can be $2$ or $3$; and if $p=7$, $D$ can be chosen to be one
of $3,$ $5$, or $6$. The reason for this is clear. The square of $x$ and of
its additive inverse $p-x$ are equal in $Z_{p}$. It then follows that
there are $\left( p-1\right) /2$ choices for $D$. This argument
fails when $p=2$, and we need to modify the methodology to handle that case.

The analysis can be illustrated in $V_{4}\left(
Z_{p}\right) $. For example, if $p=3$ a complete set of mutually unbiased
bases corresponds to the $10$ classes of commuting  spin matrices defined by
the recipe above. We represent the result in a grid whose row label 
is $j_{1}j_{2}$ and whose column
label is $k_{1}k_{2}$. The entries are $C_{a_{0}a_{1}}$.
\[
\begin{array}{cccccccccc}
& 00 & 01 & 02 & 10 & 11 & 12 & 20 & 21 & 22 \\ 
00 &  & C_{\infty } & C_{\infty} & C_{\infty} & C_{\infty} & C_{\infty} & C_{\infty} & 
C_{\infty} & C_{\infty} \\ 
01 & C_{00} & C_{10} & C_{20} & C_{02} & C_{12} & C_{22} & 
C_{01} & C_{11} & C_{21} \\ 
02 & C_{00} & C_{20} & C_{10} & C_{01} & C_{21} &  C_{11} & 
C_{02} & C_{22} & C_{12} \\ 
10 & C_{00} & C_{02} & C_{01} & C_{20} & C_{22} & C_{21} & C_{10} & C_{12} & 
 C_{11} \\ 
11 & C_{00} & C_{21} & C_{12} & C_{11} & C_{02} & C_{20} &C_{22} & C_{10} & 
C_{01} \\ 
12 & C_{00} & C_{11} & C_{22} & C_{12} & C_{20} & C_{01} & C_{21} & C_{02} & 
C_{10} \\ 
20 & C_{00} & C_{01} & C_{02} & C_{10} & C_{11} & C_{12} & C_{20} & C_{21} & 
C_{22} \\
 21 & C_{00} & C_{22} & C_{11} & C_{21} & C_{10} & C_{02} & C_{12} & C_{01} & 
C_{20} \\ 
22 & C_{00} &C_{12} & C_{21} & C_{22} & C_{01} & C_{10} & C_{11} & C_{20} & 
C_{02}
\end{array}
\]
The identity $S_{0,0}\otimes S_{0,0}$ lies in all the classes 
and each of the remaining $9^{2}-1$ tensor products is in exactly 
one class. If this grid of $81$ points is considered as a plane, then 
the set of points corresponding to two classes can be thought of as 
lines that intersect at only one 
point, the origin. This representation
gives some indication of the finite geometry implicit in the analysis. (In
particular, a set of translations of a fixed class partitions the entire
grid.) 

We used properties of finite fields to obtain the commuting classes
described in Theorem \ref{thN2Classes}, and in Appendix C we define the methodology for 
$d=p^{2}$ that generalizes to the case when $d=p^{n}$. There are two basic
ideas. The first is to use the form of the construction of the classes when $%
d=p$ but over an {\it extension} of the field $Z_{p}$, the {\it Galois} {\it %
field} $GF\left( p^{2}\right) $. This produces commuting classes $C_{\alpha }
$ of $V_{2}\left( GF\left( p^{2}\right) \right) $, where $\alpha \in
GF\left( p^{2}\right) $. The second idea is to map these classes
isomorphically to $V_{4}\left( Z_{p}\right) $ in such a way that the
symplectic product of the two-dimensional vector space over the extended
field is related to the symplectic product of the four-dimensional vector
space over the smaller field. 

\section{The MUB problem for $d=p^{n}$, $p$ prime}

The MUB problem for $d=p^{n}$ can be solved in a way similar to that 
used in the special
case treated above using suitable generalizations of the methodology. A
complication is that one cannot write down an explicit form of a function $%
f\left( x\right) $ that plays the role of $x^{2}-D$ when $n=2$ and
works in all cases when $p>2$. Instead, we must take as given $f\left(
x\right) $ with the properties summarized in Appendix D and compute it in
specific cases.

Specifically, we are guaranteed the existence of a finite field $GF\left( p^{n}\right) $ that
contains $Z_{p}$ and whose elements can be represented with the help of a
polynomial $f\left( x\right) $ of degree $n$ that is irreducible over $Z_{p}
$ and has $n$ distinct roots in $GF\left( p^{n}\right) $. The first
step is the analogue of Proposition \ref{propAppClasses}, and the proof follows the 
reasoning used in the proof of Proposition \ref{propClasses}.

Let $V_{2}\left( GF( p^{n}) \right)=\{u=(\alpha,\beta) : \alpha, \beta 
\in GF(p^{n}) \}$ and define the symplectic product:
\[
u \circ u^{\prime} \equiv \beta \alpha^{\prime}-\alpha \beta^{\prime}. 
\]
\begin{proposition}
Let $\alpha \in GF(p^{n})$ and define
subsets of the vector \\
space $V_{2}\left( GF\left( p^{n}\right) \right) $: 
\begin{eqnarray*}
C_{\alpha } &=&\left\{ \beta \left( 1,0\right) +\beta \alpha \left(
0,1\right)=\beta(1, \alpha)  :\beta \in GF\left( p^{n}\right) \right\}  \\
C_{\infty } &=&\left\{ \beta \left( 0,1\right) :\beta \in GF\left(
p^{n}\right) \right\}. 
\end{eqnarray*}
Then these are $p^{n}+1$ sets, each of which has $p^{n}$ vectors with only $%
\left( 0,0\right) $ common to any two sets. If $u$ and $v$ are in
the same set, $u\circ v=0$.
\end{proposition}

In Appendix D we provide the technical structure that justifies the
following theorem. The general argument follows the proof in the $d=p^{2}$
case, and we omit the details.

\begin{theorem}
The elements of $V_{2}\left( GF\left( p^{n}\right) \right) $ can be written
as vectors in a $2n$-dimensional
vector space over $Z_{p}$. Let
 $\left\{ e_{j}\text{, }f_{j}:0\leq %
j<n\right\}$ denote the $2n$ linearly independent vectors defined in Appendix 
D,  which satisfy
$Tr\left( e_{j}\circ f_{k}\right) =\delta
\left( j,k\right).$ The symplectic product in $%
V_{2}\left( GF\left( p^{n}\right) \right)$ is denoted by `` $\circ $ '', and $Tr$ is the trace operation.
Using indexing beginning at $0$, let $M$ denote the linear mapping that
maps $e_{j}$ to the $2n$-vector in $V_{2n}\left( Z_{p}\right) $ with a $1$
in position $2j$ and zeroes elsewhere and maps $f_{j}$ to the vector with a $%
1$ in position $2j+1$ and zeroes elsewhere. Then for every vector $u 
\in V_{2}(GF(p^{n}))$ we have $w=M(u) \in V_{2n}(Z_{p})$, and 
the symplectic products are related by 
\[
w_{1} \circ w_{2} =Tr\left( u_{1}\circ u_{2}\right). 
\]
Commuting classes of vectors $C_{\alpha }$ in $V_{2}\left( GF\left(
p^{n}\right) \right) $ map to commuting classes of vectors in $V_{2n}\left(
Z_{p}\right) $, and, consequently, define commuting classes of tensor 
products of spin matrices.
\end{theorem}

Here is the way to apply this theorem in specific cases, given $p, n$,
and an irreducible polynomial $f$ without multiple roots that generates $%
GF\left( p^{n}\right) $:

{\it Step 1}: Given a (symbolic) root $\lambda $ of 
\[f\left( \lambda \right)=\lambda^{n}+\sum_{k=0}^{n-1} c_{k}\lambda^{k}
=0,
\] find all $n$ roots in terms of $\lambda $. (If $f$ is a \textit{primitive
polynomial}, the theory guarantees that the roots have the form $\lambda
^{p^{t}}, 0 \leq t \leq n-1$.)

{\it Step 2}: Compute a set of coefficients $d_{k}\left( \lambda \right) $ from 
\[
f\left( \lambda \right) =\left( x-\lambda \right) \left(
d_{n-1}x^{n-1}+\cdots +d_{1}x +d_{0}\right) .
\]
The $d_{k}\left( \lambda \right) $ can be written as symmetric functions of
the roots and $d_{n-1}=1$.

{\it Step 3}: Compute the inverse of $f^{^{\prime} }\left( \lambda \right) $ as
an element in $GF\left( p^{n}\right) $.

{\it Step 4}: Define the bases $f_{k}=\lambda ^{k}\left( 0,1\right) $ 
and its dual $%
e_{k}=d_{k}\left( \lambda \right) \left( f^{^{\prime} }\left( \lambda \right)
\right) ^{-1}\left( 1,0\right) .$

{\it Step 5}: For each $\alpha =a_{0}+a_{1}\lambda +\ldots a_{n-1}\lambda
^{n-1}$ in $GF\left( p^{n}\right) $, express vectors in $C_{\alpha }$ as a
linear combination of the $e_{j}$'s and $f_{k}$'s with coefficients in $%
Z_{p} $: 
\[
\sum_{j=0}^{n-1}b_{j}\lambda ^{j}\left( \left( 1,0\right)
+\sum_{j=0}^{n-1}a_{j}\lambda ^{j}\left( 0,1\right) \right)
=\sum_{j=0}^{n-1}\left( x_{j}e_{j}+y_{j}f_{j}\right). 
\]

{\it Step 6}: The class corresponding to $C_{\alpha }$ and the 
corresponding set of commuting spin matrices are 
\begin{eqnarray}
C_{a_{0} \cdots a_{n-1}}&=&\left\{ \left( x_{0},y_{0},x_{1},y_{1},\ldots
,x_{n-1},y_{n-1}\right) \right\}
\nonumber \\
 S_{a_{0} \cdots a_{n-1}}&= & \{S_{x_{0},y_{0}}\otimes \cdots \otimes 
   S_{x_{n-1},y_{n-1}}\}.
\label{result}
\end{eqnarray}
The associated projections
can be computed using the methodology described in Appendix B.

To illustrate these theoretical results and the algorithm
described, we first show that the machinery used in the case $d=p^{2}$ 
is indeed a special case of the general result. Since $f\left( x\right)
=x^{2}-D=\left( x-\lambda \right) \left( x+\lambda \right) $, $d_{0}=\lambda 
$ and $d_{1}=1$. From $f^{^{\prime} }\left( \lambda \right) =2\lambda $ and $%
\left( 2\lambda \right) ^{-1}=\lambda \left( 2D\right) ^{-1}$, we have $%
e_{0}=2^{-1}\left( 1,0\right) $ and $e_{1}=\lambda \left( 2D\right)
^{-1}\left( 1,0\right) $. As usual $f_{0}=\left( 0,1\right) $ and $%
f_{1}=\lambda \left( 0,1\right) $. This is the structure used in 
Appendix C to derive Theorem \ref{thN2Classes}.

\textit{Example 1}: For two qubits, $p=n=2$, an appropriate polynomial is $f\left(
x\right) =x^{2}+x+1$. Then $f^{^{\prime} }\left( x\right) =1$. If 
$f(\lambda)=0$, then $\lambda
^{2}=\lambda +1$ is the second root, giving $d_{1}=1$ and $d_{0}=\lambda
^{2}$, since $x^{2}+x+1=(x-\lambda)( x-(\lambda +1))$.
Then 
\[
e_{0}=\lambda ^{2}\left( 1,0\right) \quad e_{1}=\left( 1,0\right) \quad
f_{0}=\left( 0,1\right) \quad f_{1}=\lambda \left( 0,1\right) .
\]
The five classes of vectors in $V_{2}\left(
GF\left( 2^{2}\right) \right) $ indexed by $\alpha=a_{0}+a_{1}\lambda$ are: 
\[
C_{0}=\{(0,0), (1,0),(\lambda ,0) ,(\lambda
^{2},0)\} =\left\{0, e_{1},e_{0}+e_{1},e_{0}\right\}.
\]
In the remaining classes we omit the $0$ vector.
\begin{eqnarray*}
C_{1}=\left\{ \left( 1,1\right) ,\left( \lambda ,\lambda \right) ,\left(
\lambda ^{2},\lambda ^{2}\right) \right\} &=&\left\{
e_{1}+f_{0},e_{0}+e_{1}+f_{1},e_{0}+f_{0}+f_{1}\right\} \\
C_{\lambda}=\left\{ \left( 1,\lambda \right) ,\left( \lambda ,\lambda
^{2}\right) ,\left( \lambda ^{2},1\right) \right\}&=&\left\{
e_{1}+f_{1},e_{0}+e_{1}+f_{0}+f_{1},e_{0}+f_{0}\right\}\\
C_{\lambda ^{2}}=\left\{ \left( 1,\lambda ^{2}\right) ,\left( \lambda
,1\right) ,\left( \lambda ^{2},\lambda \right) \right\}& =&\left\{
e_{1}+f_{0}+f_{1},e_{0}+e_{1}+f_{0},e_{0}+f_{1}\right\}\\
C_{\infty }=\left\{ \left( 0,1\right) ,\left( 0,\lambda \right) ,\left(
0,\lambda ^{2}\right) \right\}& =&\left\{ f_{1},f_{0}+f_{1},f_{0}\right\}.
\end{eqnarray*}
If one plots each of the $C_{\alpha }$ as four points in $V_{2}\left(
GF\left( 2^{2}\right) \right) $, using as coordinates the elements of $%
GF\left( 2^{2}\right) $, one obtains the left hand plots in [\cite{Woot2},
Figure 6]. The remaining plots are obtained by translation and the result is
a partition of the plane since ``parallel'' lines don't intersect.
Under the mapping $M$, 
\[
C_{0}\rightarrow C_{0,0}=\left\{ \left( 0000\right) ,\left( 0010\right)
,\left( 1010\right) ,\left( 1000\right) \right\} , 
\]
\[
C_{1}\rightarrow C_{1,0}=\left\{ \left( 0000\right) ,\left( 0110\right)
,\left( 1011\right) ,\left( 1101\right) \right\} , 
\]
\[
C_{\lambda }\rightarrow C_{0,1}=\left\{ \left( 0000\right) ,\left(
0011\right) ,\left( 1111\right) ,\left( 1100\right) \right\} , 
\]
\[
C_{\lambda ^{2}}\rightarrow C_{1,1}=\left\{ \left( 0000\right) ,\left(
0111\right) ,\left( 1110\right) ,\left( 1001\right) \right\} , 
\]
\[
C_{\infty }\rightarrow C_{\infty }=\left\{ \left( 0000\right) ,\left(
0100\right) ,\left( 0001\right) ,\left( 0101\right) \right\} , 
\]
where we abuse the notation in the last set. We can write these in terms of
the spin matrices, but it looks more familiar using Pauli matrices.
Omitting the identity $\sigma_{0}\otimes \sigma_{0}$, the classes are 
\begin{eqnarray*}
C_{0,0}\leftrightarrow \left\{ \sigma _{0}\otimes \sigma _{z},\sigma
_{z}\otimes \sigma _{z},\sigma _{z}\otimes \sigma _{0}\right\} &&
C_{1,0}\leftrightarrow \left\{ \sigma _{x}\otimes \sigma _{z},\sigma
_{z}\otimes i\sigma _{y},i\sigma _{y}\otimes \sigma _{x}\right\}\\
C_{0,1}\leftrightarrow \left\{ \sigma _{0}\otimes i\sigma _{y},i\sigma
_{y}\otimes i\sigma _{y},i\sigma _{y}\otimes \sigma _{0}\right\} &&
C_{1,1}\leftrightarrow \left\{ \sigma _{x}\otimes i\sigma _{y},i\sigma
_{y}\otimes \sigma _{z},\sigma _{z}\otimes \sigma _{x}\right\} \\ 
C_{\infty }\leftrightarrow \left\{ \sigma _{x}\otimes \sigma
_{0},\sigma _{0}\otimes \sigma _{x},\sigma _{x}\otimes \sigma 
_{x}\right\}.&&
\end{eqnarray*}
We discuss the associated projections in the next section.

\textit{Example 2}: For three qubits, $p=2$ and $n=3$, there are two primitive polynomials. 
We take $f\left( x\right) =x^{3}+x+1$. If $\lambda$ is a root, so are $\lambda ^{2}$ 
and $\lambda^{4}=\lambda +\lambda ^{2}$. 
$f^{\prime}(\lambda)=\lambda^{2}+1$ and $(\lambda^{2}+1)^{-1}=\lambda$. 
From $x^{3}+x+1=(x-\lambda)(x^{2}+\lambda x+\lambda^{2}+1)$, we get 
\[
e_{0}=\left( 1,0\right) \quad e_{1}=\lambda^{2}
\left( 1,0\right) \quad e_{2}=\lambda\left( 1,0\right). 
\]

We can summarize the subsequent analysis by writing out the classes 
$C_{a_{0}a_{1}a_{2}}$ or the sets of associated spin matrices, 
(\ref{result}). A more compact summary follows from the observation 
that each class $C_{a_{0}a_{1}a_{2}}$ is a subspace of $V_{6}(Z_{2})$ 
with a basis of three vectors defined by setting one of the $x_{j}=1$ 
and the other $x's$ to zero. The basis for $C_{\infty}$ is 
obtained by setting one of the $y_{j}=1$ and the others to zero. 
Denoting the bases by 
$G_{a_{0}a_{1}a_{2}}$ we obtain:
\begin{eqnarray*}
	 G_{000}& = & \{(100000), (001000), (000010) \}  \\
	G_{100} & = & \{(110000), (000110), (001001) \}  \\
	G_{010} & = & \{(100100), (000011), (011100) \}  \\
	G_{110} & = & \{(110100), (000111), (011101) \}  \\
	G_{001} & = & \{(100001), (010110), (001101) \}  \\
	G_{101} & = & \{(110001), (010010), (001100) \}  \\
	G_{011} & = & \{(100101), (010111), (011001) \}  \\
	G_{111} & = & \{(110101), (010011), (011000) \} \\
    G_{\infty} &=& \{(010000), (000100), (000001) \}.
\end{eqnarray*}
The spin matrices associated with the 
generators can be determined using  (\ref{result}). For example,
the set of matrices associated with the set of indices generated by $G_{010}$
is
\begin{eqnarray*}
&&\{ \sigma _{0}\otimes \sigma _{0}\otimes \sigma _{0}, \sigma _{z}\otimes
\sigma _{x}\otimes \sigma _{0},\sigma _{x}\otimes i\sigma _{y}\otimes \sigma
_{0},i\sigma _{y}\otimes \sigma _{z}\otimes \sigma _{0},\\
&&\sigma _{0}\otimes \sigma _{0}\otimes i\sigma _{y},\sigma _{z}\otimes
\sigma _{x}\otimes i\sigma _{y},\sigma _{x}\otimes i\sigma _{y}\otimes
i\sigma _{y},i\sigma _{y}\otimes \sigma _{z}\otimes i\sigma _{y}\}.
\end{eqnarray*}
Again we defer the discussion 
of the associated projectors to the next section.

\section{Separable measurements}

If $d=p^{n}$, the basic Hilbert space $H$ can be represented as an $n$-fold
 tensor product $H_{1}\otimes \cdots \otimes H_{n}$ and each factor can
be associated with a distinct subsystem. If a projection $P$ factors as $
P_{1}\otimes \cdots \otimes P_{n}$ compatible with the representation of $H$%
, then measurements can be made by coordinating local measurements at the $n$
different sites. One calls such a projection \textit{completely separable}. 
The generalization of this idea is that
\[
P=P\left( I_{1}\right) \otimes \cdots \otimes P\left( I_{m}\right) 
\]
where the $I_{k}$ are disjoint sets of indices such that $I_{1}\cup 
\cdots \cup I_{m}=\{1, \ldots, n\}$. A projection factoring this way is called $\left(
I_{1},\ldots ,I_{m}\right) $ \textit{separable}. In this case the $m$ subsystems 
can be measured separately without loss of information. If $P$ has no such factorization, we
say it is \textit{completely inseparable}. 
Separability properties of bases were discussed in some of the earlier work,
 \cite{zeil} for example. The notation here facilitates a systematic analysis.
Just as the commutativity of the spin matrices is encoded in the indices, the
nature of separability of the mutually unbiased bases is also
encoded in the indices. For example, let $n=2$ and let $p$ be odd and consider the set $%
C_{a_{0},0}=\left\{ \left( 2b_{0},a_{0}b_{0},2b_{1}D,a_{0}b_{1}\right)
\right\} $ of indices from Section 4. In the notation of Appendix B, $%
u_{1}=\left( 2,a_{0}\right) $, $u_{2}=\left( 0,0\right) $, $v_{1}=\left(
0,0\right) $, and $v_{2}=\left( 2D,a_{0}\right) $. The associated
projections computed from Appendix B are 
\begin{eqnarray*}
P_{u,v}\left( r\right)  &=&\frac{1}{p^{2}}\sum_{m_{1}}\sum_{m_{2}}\left(
\eta ^{r_{1}}S_{u_{1}}\otimes S_{0,0}\right) ^{m_{1}}\left( \eta
^{r_{2}}S_{0,0}\otimes S_{v_{2}}\right) ^{m_{2}} \\
&=&\left( \frac{1}{p}\sum_{m_{1}}\left( \eta ^{r_{1}}S_{u_{1}}\right) ^{m_{1}}\right)
\otimes  \left( \frac{1}{p}\sum_{m_{2}}\left(
\eta ^{r_{2}}S_{v_{2}}\right) ^{m_{2}}\right), 
\end{eqnarray*}
a tensor product of projections. Hence the projections associated
with $C_{a_{0},0}$ are completely separable.

The $G_{010}$ in Example 2 of Section 5 illustrates partial
separability. Using $010$ as a subscript in place of $u,v$, $P_{010}\left(
r_{1}r_{2}r_{3}\right) $ can be written as 
\[
\left( \frac{1}{4}\sum_{m_{1}}\sum_{m_{2}}\left( \left( -1\right)
^{r_{1}}\sigma _{z}\otimes \sigma _{x}\right) ^{m_{1}}\left( \left(
-1\right) ^{r_{2}}\sigma _{x}\otimes \sigma _{y}\right) ^{m_{2}}\right)
\otimes \left( \frac{1}{2}\sum_{m_{3}}\left( \left( -1\right)
^{r_{3}}\sigma _{y}\right) ^{m_{3}}\right) .
\]
We describe this as $\left( 12\right) \left( 3\right) $ separability. An
examination of the remaining cases shows that $G_{\infty }$ and $G_{000}$
are completely separable, $G_{100}$ and $G_{101}$ are $\left( 1\right)
\left( 23\right) $ and $\left( 13\right) \left( 2\right) $ separable,
respectively, and the remaining cases are completely inseparable.

These separability properties are also apparent in the basis 
vectors. For example, in Theorem 4.2 the subspace $C_{a_{0}, 0}$ of $V_{4}\left( Z_{p}\right) 
$ can be written as a direct sum of two subspaces:
\[
C_{a_{0}, 0}=span\left( \left( 2,a_{0},0,0\right) \right) \oplus span\left(
\left( 0,0,2D,a_{0}\right) \right). 
\]
In Example 2 of section 5 the subspace $C_{010}$ of $V_{6}\left( Z_{2}\right) $
can be written as 
\[
C_{010}=span\left( \left( 100100\right) ,\left( 011100\right) \right) \oplus
span\left( \left( 000011\right) \right) .
\]

The general case is the obvious extension to more indices and different
varieties of separability. We limit ourselves to a bipartite factorization
for simplicity, and we omit the proof.

\begin{theorem}
Let $I_{1}$ denote the indices of a subset of factors in $H_{1}\otimes
\cdots \otimes H_{n}$ and let $I_{2}$ denote the complementary factors. Suppose 
\[
C_{a_{0} \ldots  a_{n-1}}=C_{a_{0} \ldots  a_{n-1}}\left( I_{1}\right)
\oplus C_{a_{0} \ldots  a_{n-1}}\left( I_{2}\right), 
\]
where the vectors in $C_{a_{0} \ldots a_{n-1}}\left( I_{k}\right) $ have
zero entries in the pairs of indices not indexed by $I_{k}$. Then the
associated projections $P_{a_{0} \ldots a_{n-1}}\left( r\right) $ are $
\left( I_{1},I_{2}\right) $ separable and 
\[
P_{a_{0} \ldots a_{n-1}}=P_{a_{0} \ldots a_{n-1}}\left( r\left(
I_{1}\right) \right) \otimes P_{a_{0} \ldots a_{n-1}}\left( r\left(
I_{2}\right) \right), 
\]
where $r\left( I_{k}\right) $ has non-zero components only in positions
indexed by $I_{k}$. 
\end{theorem}

Finally, if 
\[
C_{a_{0} \ldots a_{n-1}}=\oplus _{k=1}^{m}C_{a_{0} \ldots a_{n-1}}\left(
I_{k}\right) ,
\]
then the vectors in $C_{a_{0} \ldots  a_{n-1}}\left( I_{k}\right) $ have
symplectic product zero and hence the associated spin matrices commute. 
The formal verification is easy, and we leave it to
the reader to confirm that property for the examples described above.

\section*{Acknowledgement}
We are happy to acknowledge the critical help of John Dillon and of David
Lieberman, who suggested the use of field extensions as the key methodology
and who also suggested the use of the trace operation and dual bases. 
This work was supported by NSF grants EIA-0113137 and 
DMS-0309042.

\appendix

\section{Projections of generalized spin matrices}

Here are the details for the projections associated with the $S_{u}.$ 
We recall the Definitions \ref{defAlpha} and \ref{propProj1} and prove 
Proposition \ref{propProj}.

\begin{proposition}
When $d$ is prime, $\left\{ P_{u}\left( r\right) :0\leq r<d\right\} $ is a complete set of
mutually orthogonal projections.
\end{proposition}
\textit{Proof}: We have
\[
P_{u}\left( r\right) P_{u}\left( s\right) 
=\frac{1}{d^{2}}
\sum_{m=0}^{d-1}\left( \sum_{n=0}^{d-1}\left( \alpha _{u}\right) ^{m+n}\eta
^{rm+sn}S_{u}^{m+n}\right).
\]
 Consider two cases. Suppose $0\leq n\leq
d-m-1.$ Define $t$ by $m\leq t\equiv m+n\leq d-1$ and replace this part of the $n$%
-summation by the corresponding $t$-summation. If $d-m\leq n<d$, $0\leq
t\equiv m+n-d<m$, and we have altogether 
\[
P_{u}\left( r\right) P_{u}\left( s\right) =\frac{1}{d^{2}}%
\sum_{m=0}^{d-1}\left[ \sum_{t=m}^{d-1}\eta ^{m\left( r-s\right) }\alpha
_{u}^{t}\eta ^{ts}S_{u}^{t}+\sum_{t=0}^{m-1}\eta ^{m\left( r-s\right)
}\alpha _{u}^{t+d}\eta ^{ts}S_{u}^{t+d}\right]. 
\]
Now $\alpha _{u}^{t+d}S_{u}^{t+d}=\left( \alpha _{u}S_{u}\right) ^{t}\alpha
_{u}^{d}\eta ^{jk\binom{d}{2}}$. By virtue of the definition of $\alpha _{u}$,
$\alpha _{u}^{d}\eta ^{jk\binom{d}{2}}=1$, and it is precisely for this
reason that we chose the specific form of $\alpha _{u}$. It follows that 
\[ 
P_{u}\left( r\right) P_{u}\left( s\right) = 
\frac{1}{d^{2}}\sum_{t=0}^{d-1}\left( \alpha _{u}\eta S_{u}\right)
^{t}\sum_{m=0}^{d-1}\eta ^{m\left( r-s\right) }. 
\]
When $r\neq s$, $\sum_{m=0}^{d-1}\eta ^{m\left( r-s\right) }=0$. When $r=s,$
the second summation equals $d$, and thus $P_{u}\left( r\right) P_{u}\left(
s\right) =\delta \left( r,s\right) P_{u}\left( r\right) $.

It remains to show that $\left( P_{u}\left( r\right) \right) ^{\dagger
}=P_{u}\left( r\right) ,$ and again we need $\alpha _{u}^{d}\eta ^{jk\binom{d%
}{2}}=1$. 
\begin{eqnarray*}
\left( P_{u}\left( r\right) \right) ^{\dagger } &=&\frac{1}{d}%
\sum_{m=0}^{d-1}\alpha _{u}^{-m}\eta ^{-mr}\left( \eta ^{jk\binom{m}{2}%
}S_{mj,mk}\right) ^{\dagger } \\
&=&\frac{1}{d}\sum_{m=0}^{d-1}\alpha _{u}^{-m}\eta ^{-mr}\eta ^{m^{2}jk-jk%
\binom{m}{2}}S_{-mj,-mk}
\end{eqnarray*}
where we use $m^{2}-\binom{m}{2}=\binom{%
m+1}{2}$ and the substitution $n=d-m$ for $1\leq m<d$. From the properties of
the spin matrices, we obtain 
\[
\left( P_{u}\left( r\right) \right) ^{\dagger }=\frac{1}{d}\left[
S_{0,0}+\sum_{n=1}^{d-1}\alpha _{u}^{n}\eta ^{nr}\eta ^{jk\binom{n}{2}%
}S_{nj,nk}\alpha _{u}^{-d}\eta ^{jk\binom{d}{2}}\right] =P_{u}\left(
r\right). \quad \Box
\]

\section{Projections of tensor products of generalized spin matrices}

In Theorem \ref{thN2Classes}, which solves the MUB problem for the bipartite case, 
we obtained classes of
matrices of the form $S_{2b_{0},a_{0}b_{0}+a_{1}b_{1}D}\otimes
S_{2b_{1}D,a_{0}b_{1}+a_{1}b_{0}}$ where $a_{0}$ and $a_{1}$ are fixed, and
the $b_{k}$'$s$ vary over $Z_{p}$. Following the ideas used above, we want
to show how the projections for each class can be computed from the spin
matrices in the class. From Lemma \ref{lemmaProd}
\begin{eqnarray*}
S_{2b_{0},a_{0}b_{0}+a_{1}b_{1}D} &=&S_{b_{0}\left( 2,a_{0}\right) } %
S_{b_{1}\left( 0,a_{1}D\right) } \\
S_{2b_{1}D,a_{0}b_{1}+a_{1}b_{0}} &=&S_{b_{0}\left( 0,a_{1}\right) } %
S_{b_{1}\left( 2D,a_{0}\right) }\eta ^{-b_{0}b_{1}2Da_{1}},
\end{eqnarray*}
so that, up to powers of $\eta $, matrices in this class are of the form 
\[
\left( S_{b_{0}\left( 2,a_{0}\right) }\otimes S_{b_{0}\left( 0,a_{1}\right)
}\right)  \left( S_{b_{1}\left( 0,a_{1}D\right) }\otimes S_{b_{1}\left(
2D,a_{0}\right) }\right) .
\]

Accordingly, set $u_{1}=\left( 2,a_{0}\right) $, $u_{2}=\left(
0,a_{1}D\right) $, $v_{1}=\left( 0,a_{1}\right) $, $v_{2}=\left(
2D,a_{0}\right) $. For simplicity let $u$ denote $\left( u_{1},u_{2}\right) $, 
let $v$ denote 
$\left( v_{1},v_{2}\right)$ , and let $r=\left( r_{1},r_{2}\right) $. Up to the
factor $\eta ^{-b_{0}b_{1}2Da_{1}}$ the matrices in the commuting class 
$C_{a_{0},a_{1}}$ have the form 
\[
\left( S_{b_{0}u_{1}} S_{b_{1}u_{2}}\right) \otimes \left(
S_{b_{0}v_{1}} S_{b_{1}v_{2}}\right) =\left( S_{b_{0}u_{1}}\otimes
S_{b_{0}v_{1}}\right)  \left( S_{b_{1}u_{2}}\otimes
S_{b_{1}v_{2}}\right) , 
\]
and this motivates the definition 
\[
P_{u,v}\left( r\right) \equiv \frac{1}{d^{2}}\sum_{m_{1}}\sum_{m_{2}}\left(
\eta ^{r_{1}}S_{u_{1}}\otimes S_{v_{1}}\right) ^{m_{1}} \left( \eta
^{r_{2}}S_{u_{2}}\otimes S_{v_{2}}\right) ^{m_{2}}. 
\]

\begin{proposition}
$B_{a_{0},a_{1}}=\left\{ P_{u,v}\left( r\right) :r_{1}\text{, }r_{2}\in
Z_{p}\right\} $ is the set of orthogonal projections generated by the
commuting unitary matrices indexed by $C_{a_{0},a_{1}}$.
\end{proposition}
{\it Proof}: Expand $P_{u,v}\left( r\right) $ $P_{u,v}\left( s\right) $
using $m$ and $n$ for the summation variables. Then check that
\[
\left( S_{u_{2}}\otimes S_{v_{2}}\right) ^{m_{2}} \left(
S_{u_{1}}\otimes S_{v_{1}}\right) ^{n_{1}}=\left( S_{u_{1}}\otimes
S_{v_{1}}\right) ^{n_{1}} \left( S_{u_{2}}\otimes S_{v_{2}}\right)
^{m_{2}}
\]
since $u_{1}\circ u_{2}+v_{1}\circ v_{2}=0.$ Hence, $P_{u,v}\left( r\right)
P_{u,v}\left( s\right) $ can be written as 
\[
\frac{1}{d^{4}}\sum_{k_{1}}\sum_{k_{2}}\eta ^{s_{1}k_{1}+s_{2}k_{2}}\left(
S_{u_{1}}\otimes S_{v_{1}}\right) ^{k_{1}} \left( S_{u_{2}}\otimes
S_{v_{2}}\right) ^{k_{2}}
\]
multiplied by $\sum_{m_{1}}\sum_{m_{2}}\eta ^{m_{1}\left( r_{1}-s_{1}\right)
+m_{2}\left( r_{2}-s_{2}\right) }$. It follows that the product is $
P_{u,v}\left( r\right) $ if $r=s$, and $0$ otherwise. Clearly $P_{u,v}\left(
r\right) $ has trace $1$ since only the $m_{1}=m_{2}=0$ term contributes to
the trace. We need to prove that  $P_{u,v}^{\dagger }\left( r\right)
=P_{u,v}\left( r\right).$ This can be verified using the same techniques
illustrated above and we omit the details. Finally it is easy to check that
\[
\left( S_{u_{1}}\otimes S_{v_{1}}\right) ^{t_{1}} \left(
S_{u_{2}}\otimes S_{v_{2}}\right) ^{t_{2}}=\sum_{n_{1}}\sum_{n_{2}}\eta
^{-n_{1}t_{1}-n_{2}t_{2}}P_{u,v}\left( n\right) ,
\]
 where $n=(n_{1},n_{2})$.

Analogous results can be extended to the case of multiple tensor products
using the same kind of reasoning. Since the only complication is notational,
we omit the statements and proofs.

\section{Methodology for $d=p^{2}$, $p$ an odd prime}

Anticipating step 1 of Section 5, define the polynomial $f\left( x\right) =x^{2}-D$, where $D$
is chosen so that $f(x)$ does not have a root in $Z_{p}$. Now let 
$\lambda $ denote a root of $f(x)$ in $GF(p^{2})$.
(The analogue is the introduction of the symbol $i$ to denote a root of
 $f(x)=x^{2}+1$, which does not have a root in the real
numbers.) Following \cite{Mc,Van} define the Galois field
\[
GF\left( p^{2}\right) =\left\{j+k\lambda :j,k\in Z_{p}\right\} 
\]
 with coordinate-wise addition and multiplication $\bmod$ $p$ defined by 
\begin{eqnarray*}
	 (j+k\lambda)+(a+b\lambda)& = & (j+a)+(k+b)\lambda  \\
\left( j+k\lambda \right) \left( a+b\lambda \right) &=&ja+Dkb+\lambda \left(
jb+ka\right) . 
\end{eqnarray*}
In analogy with the definition of multiplication of complex numbers, 
$\lambda^{2}=D$.
In $GF\left( p^{2}\right) $ there are two distinct solutions of $%
f\left( x\right) =0$ : $\lambda $ and $\left( p-1\right) \lambda $ where we
need $p>2$ to guarantee that these are indeed distinct elements in $GF\left(
p^{2}\right) $. The remaining exercise is to convince oneself that this
produces a field of $p^{2}$ elements.
 For example, $\left(
j-k\lambda \right) \left( j^{2}-Dk^{2}\right) ^{-1}$ is the multiplicative
inverse of $j+k\lambda $, and one sees the importance of the choice of $D$
to guarantee that $j^{2}-Dk^{2}\neq 0$.

Let $V_{2}\left( GF( p^{2}) \right)=\{u=(\alpha,\beta) : \alpha, \beta 
\in GF(p^{2}) \}$ and define the symplectic product:
\[
u \circ u^{\prime} \equiv \beta \alpha^{\prime}-\alpha \beta^{\prime}. 
\]
\begin{proposition}
Define subsets of $V_{2}\left( GF\left( p^{2}\right) \right) $ for 
each $\alpha $ in $GF\left( p^{2}\right) $
\begin{eqnarray*}
C_{\alpha }&=&\{ \beta (1,0) +\beta \alpha (0,1) = \beta(1,\alpha)
 :\beta \in GF\left( p^{2}\right) \} \\
C_{\infty }&=&\left\{
\beta \left( 0,1\right) :\beta \in GF\left( p^{2}\right) \right\}. 
\end{eqnarray*} 
Then
these are $p^{2}+1$ sets, each of which has $p^{2}$ vectors and only $%
\left( 0,0\right) $ is common to any two sets. If $u$ and $v$ are in
the same set, $u\circ v=0$.
\label{propAppClasses}
\end{proposition}
The proofs of the assertions above are exactly the same as
those in Proposition \ref{propClasses}. Although we are using a different field, the
arguments involving linear spaces are identical.

Now for the second idea. $V_{2}\left( GF\left( p^{2}\right) \right) $ is a
two-dimensional vector space over the extended field.  
$GF(p^{2})$ can be thought of as a two-dimensional space over 
$Z_{p}$.
Specifically, if $\alpha =j_{1}+j_{2}\lambda $ and $\beta
=k_{1}+k_{2}\lambda $, then $u=\left( \alpha ,\beta \right) =\alpha \left(
1,0\right) $ $+\beta \left( 0,1\right) $ can be written as 
\begin{eqnarray*}
	u &= & \left( j_{1}+j_{2}\lambda \right) \left( 1,0\right) +\left(
k_{1}+k_{2}\lambda \right) \left( 0,1\right) \\
&=&j_{1}\left( 1,0\right)
+j_{2}\lambda \left( 1,0\right) +k_{1}\left( 0,1\right) +k_{2}\lambda \left(
0,1\right) , 
\end{eqnarray*}
which motivates the representation of $V_{2}(GF(p^{2}))$ as a four-dimensional 
vector space over $Z_{p}$. However, to relate the
symplectic product in $V_{2}(GF(p^{2}))$ to the vector symplectic 
product in  (\ref{symprod1}), we take special basis vectors. Specifically,
we define 
\[
e_{0}=2^{-1}\left( 1,0\right) ,\;e_{1}=\left( 2D\right) ^{-1}\lambda \left(
1,0\right) ,\;f_{0}=\left( 0,1\right) ,\;f_{1}=\lambda \left( 0,1\right) 
\]
and use these so that 
\[
(\alpha,\beta)
=2j_{1}e_{0}+2Dj_{2}e_{1}+k_{1}f_{0}+k_{2}f_{1}. 
\]

\begin{proposition}
Let $M$ be the linear mapping from $V_{2}(GF(p^{2}))$ 
to $V_{4}\left(Z_{p}\right)$ defined by its action on $e_{r}$ and 
$f_{r}$: $M\left( e_{0}\right)
=\left( 1,0,0,0\right) ,\: M\left( e_{1}\right) =\left( 0,0,1,0\right)
,\\ M\left( f_{0}\right) =\left( 0,1,0,0\right) ,\:M\left( f_{1}\right)
=\left( 0,0,0,1\right).$
Then $M$ is a $Z_{p}$ isomorphism --- a one-to-one, onto
mapping that preserves the linear structure. Using the notation above, 
$w=M\left( \left( \alpha ,\beta \right) \right) =\left(
2j_{1},k_{1},2Dj_{2},k_{2}\right).$
\end{proposition}

We are now ready to relate the symplectic structures of $V_{2}\left(
GF\left( p^{2}\right) \right) $ and $V_{4}\left( Z_{p}\right) $. The point,
of course, is that we want to define the classes $C_{a_{0},a_{1}}$ of
Theorem 4.1 in terms of the classes $C_{\alpha }$ of Proposition 
\ref{propAppClasses}. To do
this, we need the idea of the {\it trace} of a field extension. This gets us
into the details of finite field theory, but for the specific case at
hand we can simply define it as follows. The two solutions of $f\left(
x\right) =0$ are by definition $\lambda _{1}=\lambda $ and $\lambda
_{2}=\left( p-1\right) \lambda $ , and the latter is just the additive
inverse $-\lambda $. Then define the linear function $Tr$ as follows.

\begin{definition}
$Tr\left( j+\lambda k\right) \equiv \sum_{r=1}^{2}\left( j+\lambda
_{r}k\right) =2j.$
\end{definition}

We now have all of the machinery we need for the case $d=p^{2}$.
Furthermore, the same ingredients, suitably modified, work for $d=p^{n}$.

\begin{theorem}
Let $z=(\alpha,\beta) \in V_{2}\left( GF\left( p^{2}\right) 
\right)$ and $w=M(z)$
Then 
\[
w_{1} \circ w_{2} =Tr\left( z_{1}\circ z_{2}\right) .
\]
In particular, the class $C_{\alpha }$ in $V_{2}\left( GF\left( p^{2}\right)
\right) $ maps to the class $C_{a_{0},a_{1}}$ in $V_{4}\left( Z_{p}\right) $.
\end{theorem}
{\it Proof}: If $z=\left( \alpha ,\beta \right) $ in the notation above,
then $z_{1}=2j_{1}e_{0}+2Dj_{2}e_{1}+k_{1}f_{0}+k_{2}f_{1}$. Correspondingly, let 
$z_{2}=2r_{1}e_{0}+2Dr_{2}e_{1}+s_{1}f_{0}+s_{2}f_{1}.$ We can 
compute $z_{1}\circ z_{2}$
in terms of the $e_{j}$'s and $f_{k}$'$s$. Now $e_{j}\circ e_{k}=f_{j}\circ
f_{k}=0$ and $f_{0}\circ e_{0}=2^{-1}=f_{1}\circ e_{1}$, since $\lambda
^{2}\left( 2D\right) ^{-1}=2^{-1}.$ Finally $f_{0}\circ e_{1}=\lambda 2^{-1}$
and $f_{1}\circ e_{0}=\lambda \left( 2D\right) ^{-1}.$ Since $Tr\left(
2^{-1}\right) =1$ and $Tr\left( \lambda \right) =\lambda +\left( -\lambda
\right) =0,$ we have 
\begin{eqnarray*}
Tr( z_{1}\circ z_{2}) &=&\left( k_{1}2r_{1}-2j_{1}s_{1}\right) +\left(
k_{2}2Dr_{2}-2Dj_{2}s_{2}\right)  \\
&=&\left( 2j_{1},k_{1}\right) \circ \left( 2r_{1},s_{1}\right) +\left(
2Dj_{2},k_{2}\right) \circ \left( 2Dr_{2},s_{2}\right)  \\
&=&\left( 2j_{1},k_{1},2Dj_{2},k_{2}\right) \circ \left(
2r_{1},s_{1},2Dr_{2},s_{2}\right) ,
\end{eqnarray*}
which is $w_{1} \circ w_{2} $ in $V_{4}\left(
Z_{p}\right) $ as required. $\quad \Box$

The definition of the $e$'s and $f$'s gives $Tr\left( f_{j}\circ
e_{k}\right) =\delta \left( j,k\right) $, and that was the point of defining
the weights above. All of these techniques generalize, and details are
outlined in Appendix D.

\section{Finite fields for $d=p^{n}$, $p$ prime}

We summarize the theory of finite field extensions without proofs. For
details see \cite{Mc,Van}. $GF\left( p^{n}\right) $ denotes
a finite field with $p^{n}$ elements that contains the field $Z_{p}$ as a
subfield. Up to isomorphisms, $GF\left( p^{n}\right) $ is unique and is defined using a
polynomial 
\begin{equation}
f\left( x\right) =c_{0}+\cdots +c_{n-1}x^{n-1}+x^{n}
\label{f}
\end{equation}
that is irreducible over the field $Z_{p}$. One can also assume that $f$
factors into a product $\prod_{k=1}^{n}\left( x-\lambda _{k}\right) $ with $n
$ distinct roots $\lambda _{k}$ in $GF\left( p^{n}\right) $. Using $\lambda $
to denote one of these roots, the theory guarantees that elements of $
GF\left( p^{n}\right) $ can be written as 
\[
\alpha =a_{0}+a_{1}\lambda +\cdots + a_{n-1}\lambda ^{n-1}:a_{k}\in Z_{p}.
\]
Addition in $GF\left(
p^{n}\right) $ is coordinate-wise and in multiplication, one makes 
use of $\lambda ^{n}=-\left(
c_{0}+c_{1}\lambda +\ldots +c_{n-1}\lambda ^{n-1}\right)$. 
Then the fact that 
$f\left( x\right) $ has no roots in $Z_{p}$ is used to show $GF\left(
p^{n}\right) $ is a field.

As an example, for $d=2^{2}$ it can be shown that $f\left( x\right) $ $%
=x^{2}+x+1$ is the correct polynomial, since in $Z_{p}$ $f(0)=1$ and 
$f(1)=1$. Then 
\[
GF\left( 2^{2}\right) =\left\{ 0,1,\lambda ,\lambda ^{2}=\lambda +1\right\}
. 
\]
It is easy to check that $x^{2}+x+1=\left( x+\lambda \right) \left( x+\left(
\lambda +1\right) \right) $.

Different irreducible polynomials can generate the same finite
field, but their solutions may have different properties. For example, if
$p=3$ and $n=2$, the polynomial $\tilde{f}\left( x\right) =x^{2}+2x+2$ can be used
instead of $f(x)=x^{2}-D$ with $D=2$. If $\alpha $ is a root of $\tilde{f}\left( x\right) $ in $GF\left(
3^{2}\right) $, then  $\lambda =\alpha ^{2}$ is a root of  $f\left(
\lambda \right) =\lambda ^{2}-2.$ As an exercise in the notation, one can
confirm that $\alpha $ is a {\it primitive} root in the sense that all of
the non-zero elements of $GF\left( 3^{2}\right) $ can be written as powers
of $\alpha $. The theory guarantees primitive polynomials for finite fields,
but we do not assume any properties of the generating
irreducible polynomials beyond those set forth in the first
paragraph of this section.

The {\it trace} operation generalizes in the following way.

\begin{definition}
For each $\alpha =\alpha \left( \lambda \right) =a_{0}+a_{1}\lambda 
+\cdots
+a_{n-1}\lambda ^{n-1}$, 
\[
Tr\left( \alpha \right) \equiv \sum_{r=1}^{n}\alpha \left( \lambda
_{r}\right), 
\]
where the $\lambda _{r}$ are the distinct roots of $f\left( x\right) $ in $%
GF\left( p^{n}\right) $.
\end{definition}

For example, take $GF\left( 2^{2}\right) $. Then $Tr\left( 1\right) =0$,
$Tr\left( \lambda \right) = \lambda +\left( \lambda +1\right)
=1$, and $Tr\left( \lambda +1\right)=1 .$

From the representation of elements of $GF\left( p^{n}\right) $,
 $GF\left( p^{n}\right) $ can be considered as an $n$
dimensional space over $Z_{p}.$ Then $V_{2}\left( GF\left( p^{n}\right)
\right) $ can be written as a $2n$%
-dimensional space over $Z_{p}.$ We define $n$ of the basis vectors as
 $f_{k}=\lambda ^{k}\left( 0,1\right), $
$0\leq k\leq n-1$, as before, and we want a {\it dual} {\it basis}
consisting of vectors 
\[
\left\{ e_{j}=g_{j}\left( \lambda \right) \left( 1,0\right) :0\leq j\leq
n-1\right\} 
\]
that are linearly independent over $Z_{p}$ and satisfy 
\[
Tr\left( e_{j}\circ f_{k}\right) =Tr\left( g_{j}\left( \lambda \right)
\lambda ^{k}\right) =\sum_{r=1}^{n}g_{j}\left( \lambda _{r}\right) \lambda
_{r}^{k}=\delta \left( j,k\right) . 
\]
The remainder of this Appendix is devoted to deriving the form of $%
g_{j}\left( \lambda \right).$ Examples in Section 5 illustrate the use of
this machinery, and we follow the presentation in \cite{Lange}. For an 
alternative method to compute the dual basis based on 
primitive polynomials see \cite{Mc}.

Since $f\left( x\right) $ does not have multiple roots, $f\left( x\right) $
and $f^{^{\prime} }\left( x\right) $ have no common non-constant factors and,
in addition, $f^{^{\prime} }\left( \lambda \right) \neq 0$. From $f\left(
x\right) =\prod_{j=1}^{n}\left( x-\lambda _{j}\right)$,  $f^{^{\prime} }\left(
\lambda _{r}\right) =\prod_{j\neq r}\left( \lambda _{r}-\lambda _{j}\right) $%
. With $\lambda $ denoting a generic root, one can check that there are
values $d_{k}=d_{k}\left( \lambda \right) $ such that 
\[
\frac{f\left( x\right) }{x-\lambda }=d_{0}+d_{1}x+\cdots +d_{n-1}x^{n-1}.
\]
Combining these results, we define 
\[
F_{k}\left( x\right) \equiv \sum_{r=1}^{n}\frac{f\left( x\right) }{x-\lambda
_{r}}\frac{\lambda _{r}^{k}}{f^{^{\prime} }\left( \lambda _{r}\right) }%
=\sum_{j=0}^{n-1}x^{j}\sum_{r=1}^{n}\frac{d_{j}\left( \lambda _{r}\right) }{%
f^{^{\prime} }\left( \lambda _{r}\right) }\lambda _{r}^{k}.
\]

Now if we set $\lambda =$ $\lambda _{t}$ for each of the $n$ distinct roots,
only the $r=t$ term survives in the middle expression, so that $F_{k}\left(
\lambda _{t}\right) =\lambda _{t}^{k}$. By the general theory of polynomials
over finite fields $F_{k}\left( x\right) $ must then equal $x^{k}$. Thus 
\[
\delta \left( j,k\right) =\sum_{r=1}^{n}\frac{d_{j}\left( \lambda
_{r}\right) }{f^{^{\prime} }\left( \lambda _{r}\right) }\lambda
_{r}^{k}=Tr\left( \frac{d_{j}\left( \lambda \right) }{f^{^{\prime} }\left(
\lambda \right) }\lambda ^{k}\right),
\]
 and we have a key result.

\begin{proposition}
If $e_{j}=g_{j}\left( \lambda \right) \left( 1,0\right) $, where $%
g_{j}\left( \lambda \right) =d_{j}\left( \lambda \right)/f^{^{\prime}
}\left( \lambda \right) $, and $f_{k}=\lambda ^{k}\left( 0,1\right) $, then 
\[
Tr\left[ f_{k}\circ e_{j}\right] =\delta \left( j,r\right),
\]
and the set $\{e_{j}, f_{k}\}$ is linearly independent over $Z_{p}$.
\end{proposition}

It remains to show how to compute $d_{j}(\lambda)$. From  (\ref{f}) 
and $f\left( x\right)=\left( x-\lambda \right)
(d_{0}+d_{1}x+\cdots +d_{n-1}x^{n-1})$, $d_{n-1}=c_{n}=1$. It follows 
for $1\leq r\leq n$ that
\[
d_{n-r}=\sum_{j=0}^{r-1}\lambda ^{j}c_{n+j+1-r}.
\]
The highest order term of $d_{n-r}$ is $\lambda ^{r-1}$.

\smallskip

\end{document}